# Deep-learning-based Hyperspectral imaging through a RGB camera

Xinyu Gao,[1] Tianliang Wang,[1] Jing Yang,[1] Jinchao Tao,[1] Yanqing Qiu,[1] Yanlong Meng,[1] Bangning Mao,[1] Pengwei Zhou,[1] and Yi Li[1,*]

[1]*College of Optical and Electronic Technology, China Jiliang University, Hangzhou 310018, China*
*\*Corresponding author: yli@cjlu.edu.cn*

**Abstract:** Hyperspectral image (HSI) contains both spatial pattern and spectral information which has been widely used in food safety, remote sensing, and medical detection. However, the acquisition of hyperspectral images is usually costly due to the complicated apparatus for the acquisition of optical spectrum. Recently, it has been reported that HSI can be reconstructed from single RGB image using convolution neural network (CNN) algorithms. Compared with the traditional hyperspectral cameras, the method based on CNN algorithms is simple, portable and low cost. In this study, we focused on the influence of the RGB camera spectral sensitivity (CSS) on the HSI. A Xenon lamp incorporated with a monochromator were used as the standard light source to calibrate the CSS. And the experimental results show that the CSS plays a significant role in the reconstruction accuracy of an HSI. In addition, we proposed a new HSI reconstruction network where the dimensional structure of the original hyperspectral datacube was modified by 3D matrix transpose to improve the reconstruction accuracy.

**Keywords:** Hyperspectral Resconstruction, Deeplearning, Camera Spectral Sensitivity, Computational Imaging, Image Processing

## 1. Introduction

Hyperspectral imaging is a technique combining spectroscopy and imaging. For each pixel, many narrow band spectra can be acquired associated with the spatial information. The spectral information provides a unique way to identify materials based on their own specific spectral features. Therefore, HSI finds a wide range of applications in remote sensing [1-4], medical detection [5], agricultures [6, 7], food safety [8, 9] and many other fields [10-12]. Compared with normal cameras, the hyperspectral cameras are often equipped with special designed optical apparatus, such as optical gratings, prisms, filters and other spectroscopic components [13-26].They are of large size, complex structure and high-cost. These shortcomings limit the applications of hyperspectral imaging technique, especially in the measurements demanding low-cost and quick deployment. Recently, reconstruction of HSI from single RGB image provides a simple and low-cost solution to address the above problems. For normal cameras, the spectral information is mapped to RGB values through the camera filters. In principle, it is possible to recover the spectrum from these RGB values. But it is a severe ill-posed problem because part of spectral information is lost in the construction of the RGB image. The reconstruction algorithm is required to correctly build the non-linear mapping relationship between RGB values and the corresponding spectrum.

As for the algorithms, researchers focused on building sparse coding from specific HSI at the beginning. Arad *et al*. proposed a method to reconstruct hyperspectral images from RGB images by using sparse dictionary. The sparse dictionary was created based on hyperspectral prior [27]. In another study, Yan *et.al* proposed a manifold-based reconstruction pipeline to map an RGB vector to its corresponding hyperspectral vector [28]. In recent years, with the extensive success of deep learning in the field of image processing, CNN has been widely used in HSI reconstruction. Yan *et al.* also reported a super-resolution HSI from single RGB Image by using multi-scale CNN [29], where a symmetrical encoder-decoder structure was proposed to extract spectral features and reconstruct hyperspectral images. Shi *et al.* proposed two HSI recovery algorithms, HSCNN-R and HSCNN-D [30]. HSCNN-R is a deep residual network [31] where a large number of residual blocks were used to greatly improve the accuracy of the reconstructed HSIs. In the HSCNN-D, the residual blocks are replaced by the dense blocks with a novel fusion scheme [32], further improving the reconstruction accuracy. Very recently, Li *et al.* proposed the AWAN algorithm [33], where high-value information in the spatial and wavelength domains are picked out through attention mechanism [34].

In the employment of the above algorithms, public hyperspectral datasets were used to reconstruct the HSIs. However, in practice, the cameras could be quite different in the hardware, especially in the CCD or CMOS filters. The difference in the filters will result in a different response to the same spectrum. In this study, we built a tunable narrow band light source to calibrate the camera filter. And the experimental results show that the CSS has a significant influence on the accuracy of the final reconstructed HSI. Inspired by the 3D convolution, we also proposed a new network where the dimensional structure of the original hyperspectral datacube was modified by 3D transpose. Since the three datacubes with different dimensional structure are used as the input of three parallel sub-modules, the newly proposed network is named Parallel Transposed Network (PTNet). An RGB camera was modified to a high-speed and accurate hyperspectral camera with this algorithm.

## 2. Methods

For supervised learning algorithms, both RGB images and their corresponding HSIs are essential for network training. However, in the practical collection of training datasets, it is difficult to take the completely same scene using the RGB camera and the HSC respectively. In order to address the above problem, computational imaging technique was introduced to generate virtual RGB images, which are approximate to the real ones. Mathematically, the imaging process of an RGB camera can be regarded as the multiplication of the object's spectrum and the CSS.

$$I_{RGB}(x,y,3) = HSI(x,y,n_\lambda) \times CSS(n_\lambda,3) \qquad (1)$$

In Eq. 1, $I_{RGB}$ represents the gray scales of the Red, Green, and Blue channels. $n_\lambda$ is the number of spectral bands, which is the same in CSS and HSI. Using this equation, we can conveniently create a training RGB dataset based on HSIs and the CSS.

*2.1 Camera spectral sensitivity*

It is well known that CSS is directly determined by its color filter. In order to ensure the generated RGB images as close as possible to the real taken ones, the CSS should be calibrated accurately. In this study, we used a Xe and a monochromator to obtain the CSS by recording the response of the camera at different wavelengths. It should be noted that the imaging process of a RGB camera usually includes automatic white balance (AWB), which is harmful to the subsequent network training. Because the AWB adjusts RGB values separately, the mapping relationship between RGB values and the corresponding spectra becomes uncertain. To solve this problem, we fixed the RGB gain to 1:1:1 in calibrating the CSS. In addition to AWB, exposure time is another factor affecting RGB values. Therefore, the exposure time should be constant in the calibration of CSS. In CNN based hyperspectral reconstruction algorithm, the taken HSIs were used as labels to compare with the predicted ones, and to calculate the loss. Then, the backward algorithm updates the weights according to the loss.

*2.2 HSI reconstruction algorithm*

HSI usually has three dimensions, two spatial dimensions (W and H) and one wavelength dimension($\lambda$). All three dimensions are equally important for HSIs. In most state-of-the-art hyperspectral reconstruction algorithms, only the wavelength dimension is taken as channel in 2D convolution operations. The network width corresponds to the number of channels. A wider network can enable each layer of the network to learn more abundant features. Thus, these algorithms increase the network width for improving the spectral reconstruction effect. However, the increase of the network width will lead to the heavy burden in GPU memory and computation. To address this problem, we proposed the PTNet for training the hyperspectral data. In this network, the original datacube is transposed to three datacubes with different dimensional structures. As shown in Fig.1, the original datacube W×H×λ is transposed into W×λ×H and λ×H×W through 3D matrix transpose. The red cube represents the hyperspectral datacube, and the blue arrow represents the 3D matrix transposition. The structure of PTNet is shown in Fig.2, which can be divided into three stages. The first stage includes an input block and a down sampling block. The role of the former is channel expansion so as to allow each layer to learn richer features, and the latter is to reduce the spatial resolution of the datacube, thereby reducing the computational burden. The second stage contains 3 residual attention blocks (RA_Block) [35]. Their inputs are the original datacube and two transposed datacubes. The transposed datacubes come from the original datacube after downsampling. The outputs of RA_Blocks are transposed and interpolated to the same shape, and the three are concatenated according to the wavelength dimension through the concatenate layer to obtain a new datacube. The third stage contains one batch normalization layer and output block, and the original datacube is added as the residual part to the output of the batch normalization layer. Since the inputs of the three blocks in stage 2 are not affected by the inputs or outputs of the other two, the network is called a parallel transposed network.

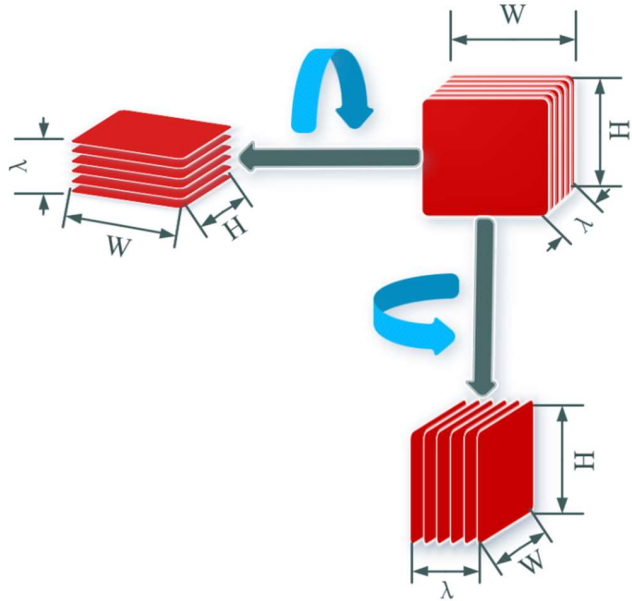

Fig 1 The schematic of transposing the hyperspectral matrix

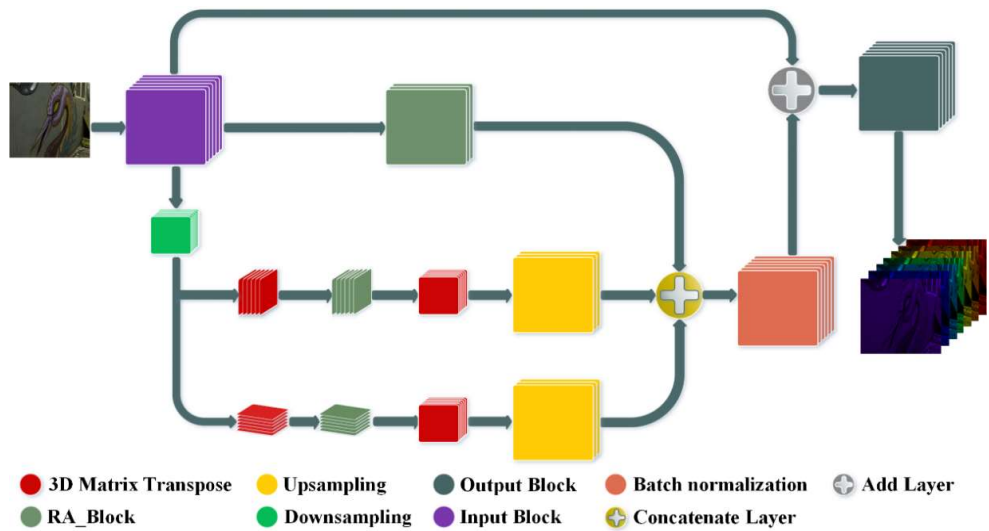

Fig 2 The diagram of the proposed PTNet

## 3. Experiments

### *3.1 camera spectral sensitivity*

As mentioned above, the CSS depends on the camera's filter features. So different cameras will produce different images even if they take shot of the same object. And the mapping relationship between spectrum and RGB value varies with different cameras. Therefore, CSS should be calibrated before using cameras for hyperspectral reconstruction. In the following experiments, we measured the CSS of a camera (Mindvision mv-sua500c-t) used in this study by a monochromator (Zolix Omni-λ3027i).

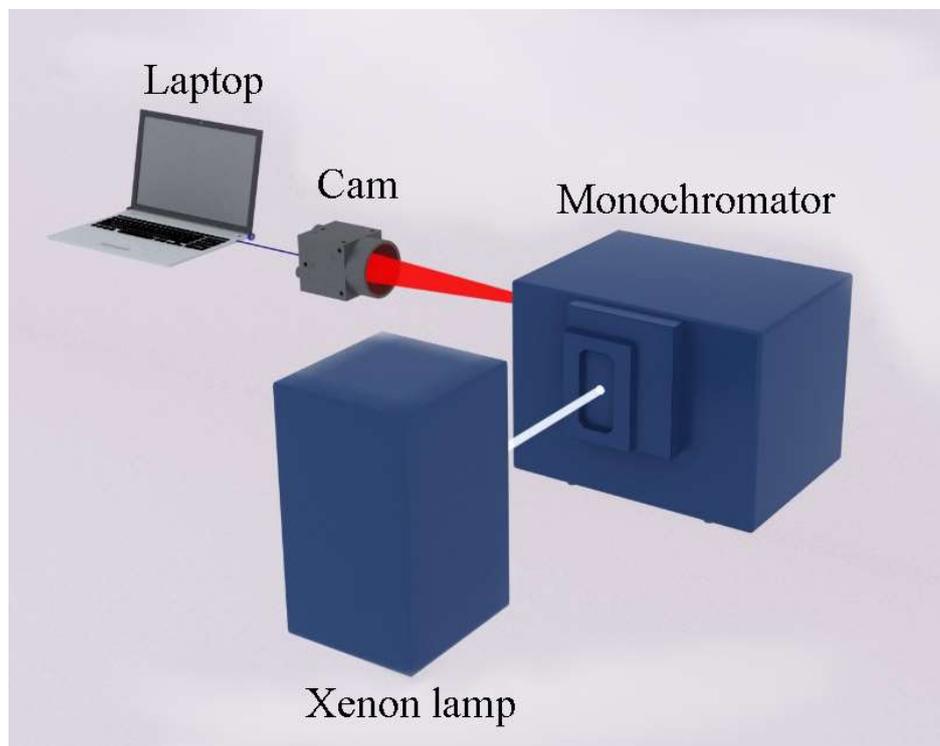

**Fig 3** The experimental setup for measuring CSS

As shown in Fig. 3, light emitted from a Xenon lamp was first passed into the monochromator. Then the narrow band light from the monochromator was tuned from 400nm to 650nm with the step of 1nm, while the camera took pictures simultaneously. During the calibration, the RGB gain and the exposure time were set to 1:1:1 and 20ms respectively. Thus, the CSS curves of each pixel can be obtained by plotting the RGB values with the corresponding wavelengths. The Fig. 4 is the typical CSS curves of one pixel. Here, the CSS curves have been compensated by the standard spectrum of the Xenon lamp.

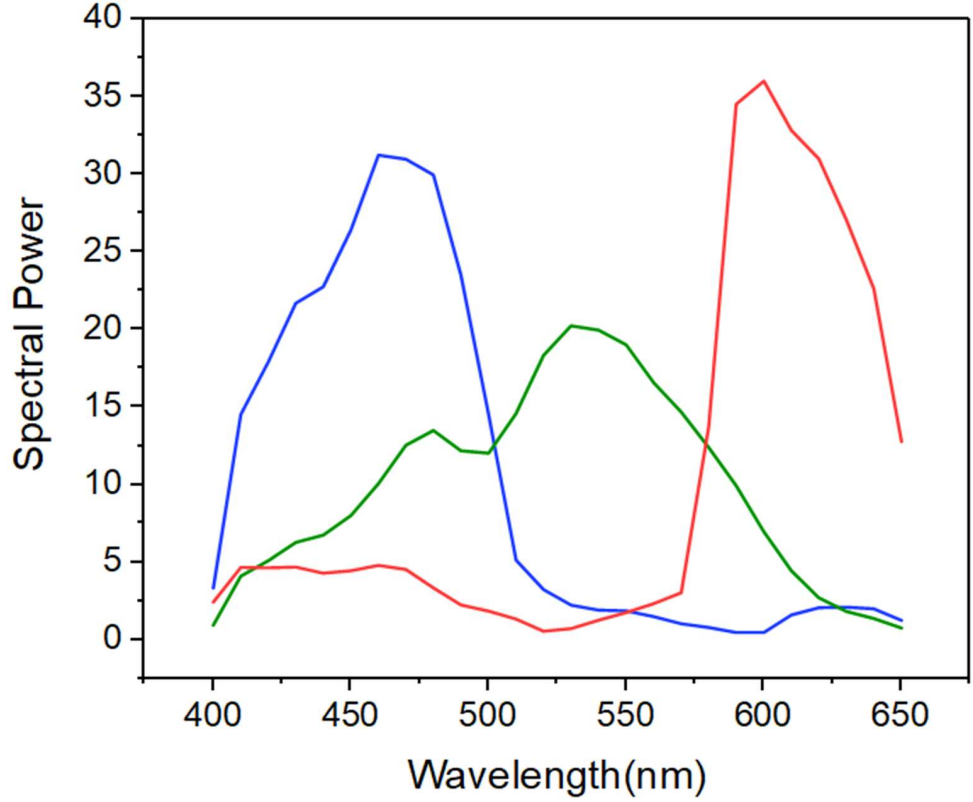

**Fig 4** The measured CSS curves of our camera

*3.2 Network Setting and Training*

In this study, the public HSI datasets from NTIRE2020 [36] were introduced to train the network, where 450 normalized images were used for training and another 10 were used for validation. The spatial resolution of the original image is 482×512. And, each image was divided to 8 patches with a spatial resolution of 241×128, for saving the GPU memory. After comparing various loss functions, a loss function based on Hamming distance had smaller reconstruction error. This loss function takes the logarithm of the Hamming distance between predicted HSI ($HSI_p$) and real HSI ($HSI_{GT}$). The Hamming distance detects the different numbers of corresponding bits in the two groups of binary data. This loss function is used to evaluate the difference between the $HSI_p$ and the corresponding $HSI_{GT}$, as shown below.

$$Loss = \alpha \cdot \log_{10} \| HSI_p, HSI_{GT} \|_{dis} \qquad (2)$$

In Eq. 2, $\|\cdot\|_{dis}$ is Hamming distance. α is the coefficient.

The detailed network training implementations are shown below. The proposed PTNet has been implemented in the Pytorch framework on a GeForce RTX 3080 GPU. The batch size of our model is 2 and the optimizer is the Ranger [37] with betas = (0.9, 0.999), eps = $10^{-8}$ and weight_decay = $10^{-3}$. The learning rate (LR) is initialized as $6\times10^{-4}$, changing as the cosine law. After some epochs, the LR drops to 0, and then back to a high value. We stopped the network training at 300 epochs, which costing about 100 hours.

## 4. Result and Discussion

The trained model was evaluated by using MRAE (Mean relative absolute error) and RMSE (Root Mean Squared Error), as shown below.

$$\mathrm{MRAE} = \frac{1}{N}\sum_{m=1}^{N}(|HSI_p^{(m)} - HSI_{GT}^{(m)}|)/HSI_{GT}^{(m)} \quad (3)$$

$$\mathrm{RMSE} = \sqrt{\frac{1}{N}\sum_{m=1}^{N}(|HSI_p^{(m)} - HSI_{GT}^{(m)}|)^2} \quad (4)$$

By comparing the reconstructed HSIs via the trained model with the $HSI_{GT}$s, MRAE=0.0354 and RMSE=0.0145 of the entire validation set were obtained. The error comparison with the other three algorithms is shown in Fig.5. We selected five bands of an image in the validation set to evaluate the reconstruction quality. For each algorithm, the reconstruction error was relatively large in the range of 650nm-700nm. The possible reason is that the CIE1964 has a weak response in this wavelength range. Apparently, PTNet has better performance in most bands, especially the long wavelength band according to Fig.6. To achieve the results, only 5.97M of the model parameters are needed. And the reconstruction of a single image takes about 7 seconds. However, the MRAE is 0.0410 when only one block is reserved. And the quality of reconstructed HSIs is much lower than using the complete PTNet. Obviously, the PTNet sufficiently extracts information from the same datacube in different dimensions arrangement, and the reconstructed hyperspectral image is more accurate.

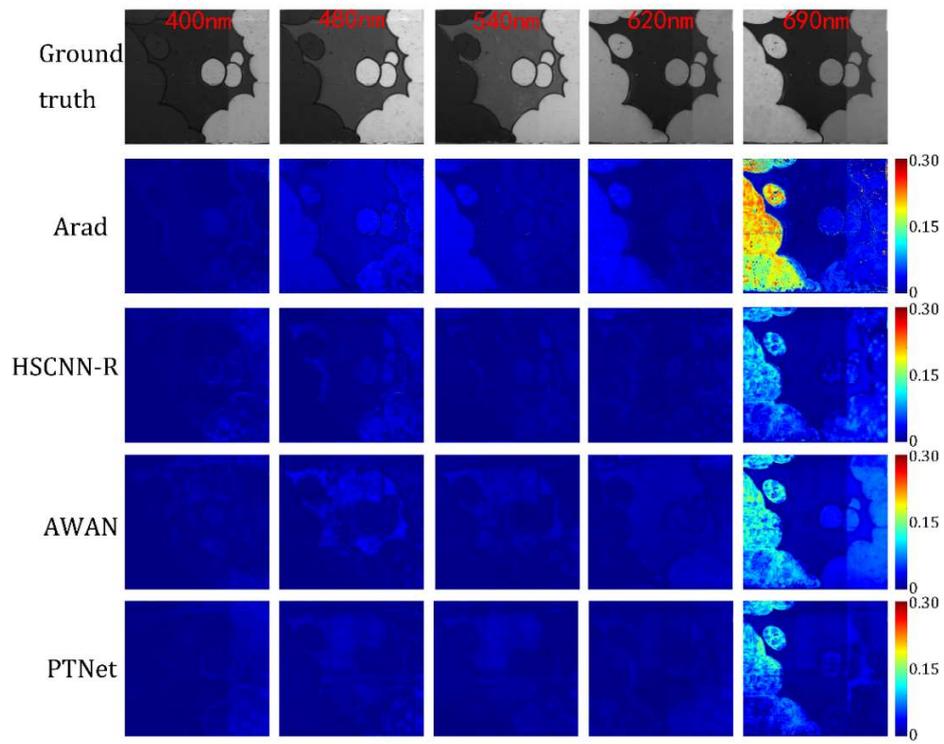

**Fig 5** Heat map comparison among different algorithms at five selected bands

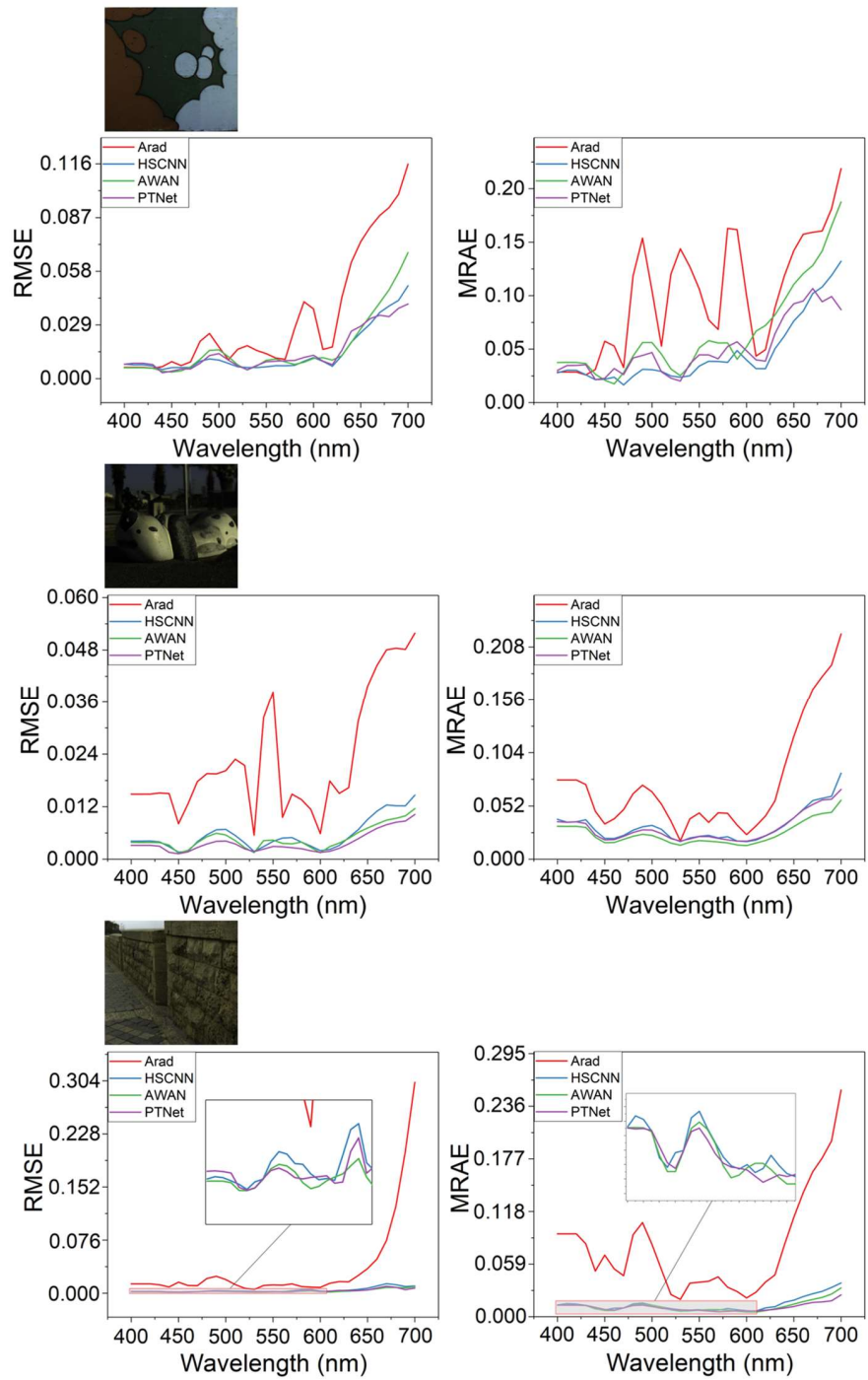

**Fig 6** Quantitative comparisons using MRAE and RMSE

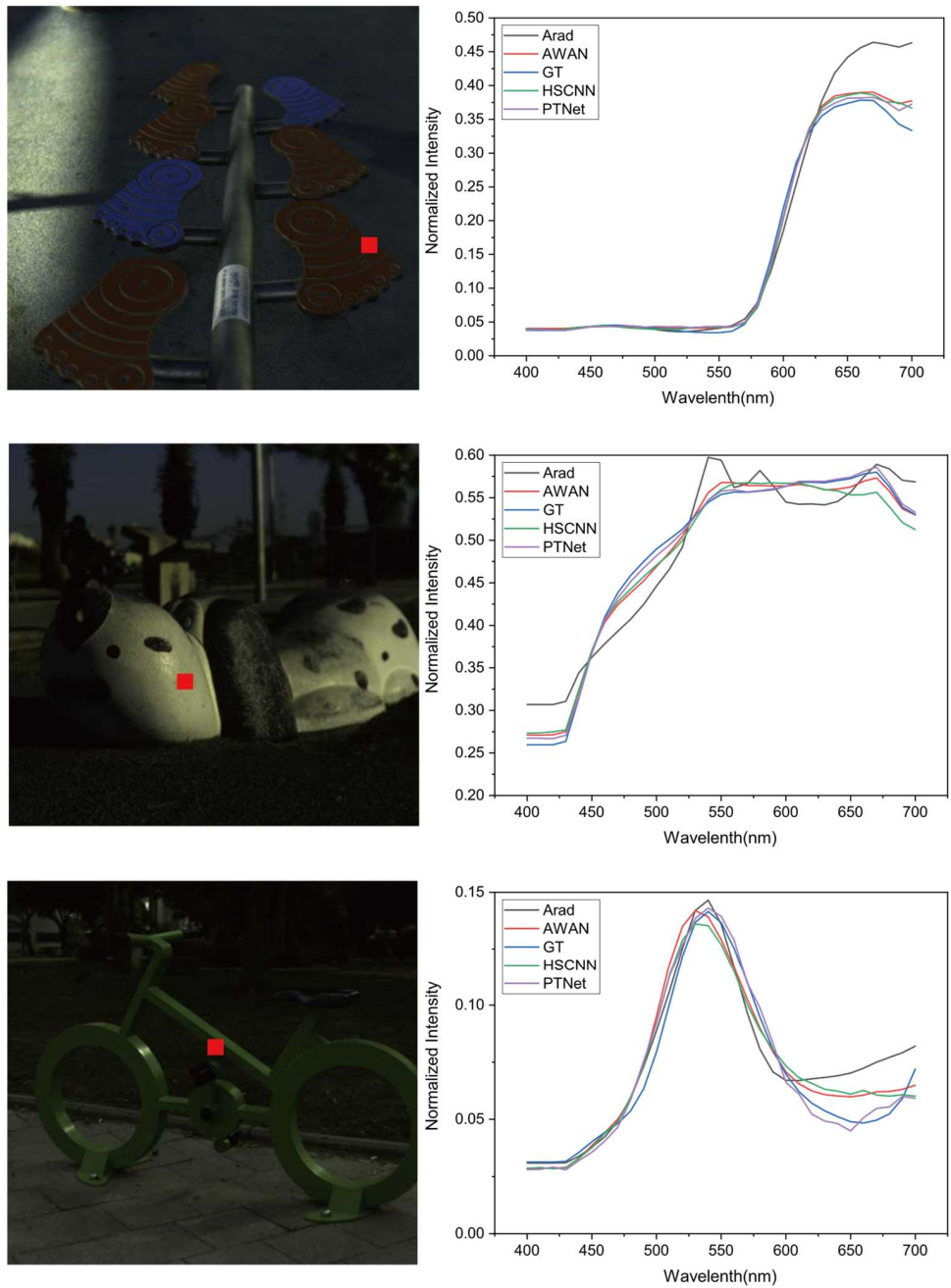

**Fig 7** Spectral comparison of selected points in hyperspectral images (Comparison of 4 kinds of hyperspectral reconstruction algorithms and the ground truth (GT))

In addition, we also investigated the influence of CSS on the versatility of the trained model. We used a CSS to generate RGB images for training with the corresponding HSIs. The trained model reconstructs three groups of RGB images generated by different CSS to HSIs.

The CSS used to generate the training set comes from our own camera. The three groups of validation images were generated based on CIE1964, D40 (public CSS of the NTIRE2020) and the measured CSS, as the following. Obviously, the mapping relationship between RGB images and their corresponding HSIs is different due to the CSS diversity. Thus, their trained model cannot be universal. Therefore, the determination of CSS is a necessary step for achieving hyperspectral imaging of real RGB cameras.

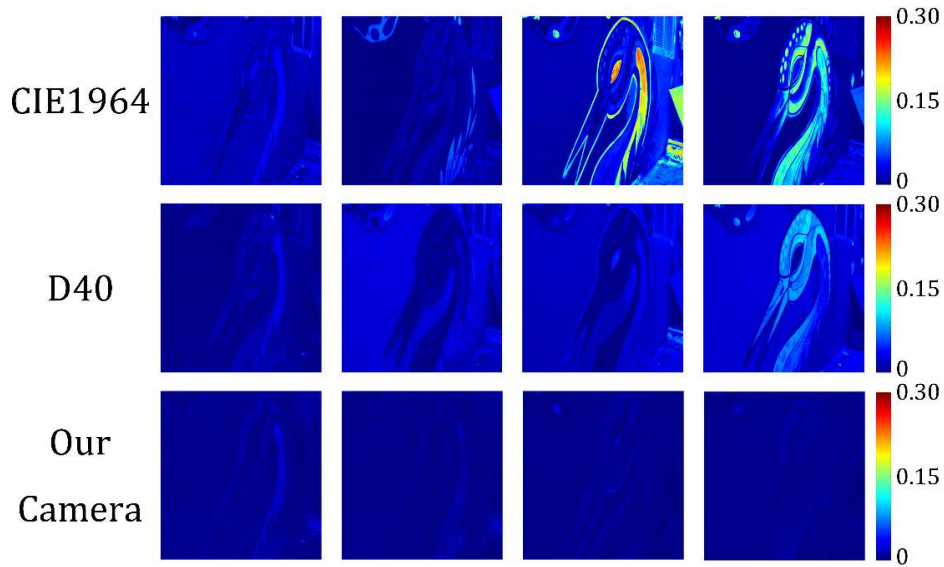

Fig 8 The heat map comparison for reconstruction error of validation sets based on different CSS

## 5. Conclusion

In this work, we reported a CNN based method to reconstruct HSI from a single RGB image according to the CSS of a real RGB camera. Compared with the traditional hyperspectral camera, our method has the advantages of low cost, easy implementation and portability. Instead of processing the public data, we explored the influence of camera hardware diversity on hyperspectral reconstruction. A Xenon lamp incorporated with a monochromator was used to measure the CSS precisely. Experimental result shows that accurate CSS helps to reduce the reconstruction error in hyperspectral imaging. Furthermore, we proposed a new reconstruction algorithm PTNet, which can deeply extract the relationship between spectrum and spatial pattern from datacubes by changing their dimensional structure. This operation improved the accuracy while ensuring the reconstruction speed. The reconstruction of a single image takes about 0.45 seconds under our experimental conditions.

**Funding.** Zhejiang Xinmiao Talents Program (2020R409042); National Natural Science Foundation of China (62075202); Natural Science Foundation of Zhejiang Province (LY20F050008).

**Disclosures.** The authors declare no conflicts of interest.